# Nucleon-Nucleon Correlations and Multiquark Cluster Effects in Deep Inelastic Electron Scattering off Few-Nucleon Systems at $x > 1$

Silvano Simula *

Istituto Nazionale di Fisica Nucleare, Sezione Sanitá, Viale Regina Elena 299, I-00161 Roma, Italy

**Abstract.** Inclusive $A(e, e')X$ and semi-inclusive $A(e, e'N)X$ deep inelastic electron scattering processes off few-nucleon systems are investigated at $x > 1$, showing some of the relevant features of the cross section which are sensitive to the effects arising from nucleon-nucleon correlations and possible exotic multiquark cluster configurations at short internucleon separations.

The aim of this contribution is to address few relevant questions concerning inclusive $A(e, e')X$ and semi-inclusive $A(e, e'N)X$ deep inelastic scattering ($DIS$) of electrons off few-nucleon systems for values of the Bjorken variable $x = Q^2/2M\nu > 1$ (corresponding to kinematical regions forbidden on a free nucleon), assuming that virtual photon absorption occurs on a hadronic cluster which can be either a nucleon-nucleon ($NN$) correlated pair or a six-quark ($6q$) bag. The cross section for the inclusive process $A(e, e')X$ can be written as

$$\sigma^A \equiv d^2\sigma/dE_{e'} \ d\Omega_{e'} = \sigma_0^A + \sigma_1^A \qquad (1)$$

where the contributions from different final nuclear states have been explicitly separated out, namely $\sigma_0^A$ describes the transition to the ground and one-hole states of the (A-1)-nucleon system and $\sigma_1^A$ the transition to more complex excited configurations (mainly 1p-2h states) arising from 2p-2h excitations generated in the target ground-state by $NN$ short-range and tensor correlations. In what follows, the $DIS$ contribution to $\sigma_0^A$ and $\sigma_1^A$ will be calculated within the impulse approximation ($IA$). As is well known, the $IA$ calculation requires the knowledge of the nucleon spectral function $P^N(k, E)$, which represents the joint probability to find in a nucleus a nucleon with momentum $k \equiv |\mathbf{k}|$ and removal energy $E$. In presence of ground-state $NN$ correlations the spectral function can be written as $P^N(k,E) = P_0^N(k,E) + P_1^N(k,E)$, where the indeces 0 and 1 have the same meaning as in Eq. (1). For $A = 3$ and 4 one gets

$$P_0^N(k,E) = n_0(k)\ \delta(E - E_{min}) \qquad (2)$$

where $n_0(k)$ is the nucleon momentum distribution corresponding to the ground-to-ground transition and $E_{min} = 5.5,\ 20\ MeV$, respectively. As for $P_1^N$, its momentum and removal energy structure, generated by $NN$ correlations, can be accounted for by adopting the extended two-nucleon correlation ($2NC$) model of Ref. [1], viz.

---

*E-mail address: simula@hpteo2.iss.infn.it

$$P_1^{N_1}(k,E) = \sum_{N_2=n,p} \int dk_{CM}\ n_{rel}^{N_1 N_2}(k - \frac{k_{CM}}{2})\ n_{CM}^{N_1 N_2}(k_{CM}) \cdot$$
$$\delta[E - E_{thr}^{(2)} - \frac{A-2}{2M(A-1)}(k - \frac{A-1}{A-2}\ k_{CM})^2] \qquad (3)$$

where $n_{rel}^{N_1 N_2}$ ($n_{CM}^{N_1 N_2}$) is the momentum distribution of the relative (center-of-mass ($CM$)) motion of the two nucleons in a correlated pair, and $E_{thr}^{(2)} = M_{A-2} + 2M - M_A$ is the two-nucleon break-up threshold. For $A = 2$ one has $P_1^N = 0$ and $P_0^N(k,E) = n(k)\ \delta(E - E_{min})$, where $n(k)$ is the nucleon momentum distribution in the deuteron and $E_{min} = 2.226\ MeV$ is the deuteron binding energy. Then, the contribution $\sigma_\alpha^{A(3q)}$ ($\alpha = 0,\ 1$), due to processes in which the struck quark belongs to a nucleon in the nucleus $A$, reads as follows

$$\sigma_\alpha^{A(3q)} = \sigma_{Mott} \sum_{N=1}^{A} \int dW dk dE\ \delta\left[\nu + k^0 - \sqrt{W^2 + (k+q)^2}\right] \cdot$$
$$P_\alpha^N(k,E)\ \frac{M}{\sqrt{M^2 + k^2}}\ \frac{W}{\sqrt{W^2 + (k+q)^2}} \sum_{j=1,2} b_j^N\ W_j^N(W,Q^2) \qquad (4)$$

where $k^0 \equiv M_A - \sqrt{(M_A + E - M)^2 + k^2}$ is the initial nucleon energy in the lab system and $W_j^N$ is the nucleon structure function. In what follows, the free nucleon structure function, parametrized as in [2], will be adopted. In Eq. (4) $b_1^N = a_{21} + 2tg^2(\theta_e/2)\ a_{11}$ and $b_2^N = a_{22} + 2tg^2(\theta_e/2)\ a_{12}$, where the explicit expressions of the coefficients $a_{ij}$ can be found in [3].

Let us now consider the possibility that the two nucleons in a correlated pair can loose their identity at short separations, so that the incoming photon can interact with a $6q$ bag structure. This is the mechanism proposed and applied to the investigation of inclusive $DIS$ processes off the deuteron in Ref. [4]. Within such a multiquark cluster picture, the inclusive cross section for the deuteron can be written as the incoherent sum of the contributions resulting from virtual photon absorption by a $3q$ and $6q$ clusters [4], viz.

$$\sigma_0^{2H} = P_{3q}^{2H}\ \sigma_0^{2H(3q)} + P_{6q}^{2H}\ \sigma^{2H(6q)} \qquad , \qquad \sigma_1^{2H} = 0 \qquad (5)$$

where $P_{3q}^{2H}$ and $P_{6q}^{2H}$ are the probability that the struck quark belongs to a nucleon of a $NN$ pair or to a $6q$ bag in the deuteron, respectively ($P_{3q}^{2H} + P_{6q}^{2H} =$

1). In Eq. (5) the quantity $\sigma^{^2H(6q)}$, representing the contribution of the process in which the struck quark belongs to a $6q$ bag, is given by

$$\sigma^{^2H(6q)} = \sigma_{Mott} \{W_2^{(6q)}(W_{6q}^*, Q^2) + 2tg^2(\theta_e/2)\ W_1^{(6q)}(W_{6q}^*, Q^2)\} \qquad (6)$$

where $W_{6q}^* \equiv \sqrt{(\nu + 2M)^2 - |q|^2}$ is the invariant mass produced by virtual photon absorption a $6q$ bag at rest (with mass $2M$) and $W_i^{(6q)}$ is the inclusive structure function of a $6q$ bag. In what follows, $W_i^{(6q)}$ will be parametrized according to the the prescriptions of Ref. [4] based on quark counting rules. For $A > 2$ only the correlated part $\sigma_1^A$ can be affected by the presence of $6q$ bag configurations, for $\sigma_0^A$ is related only to final states of the residual $(A-1)$ system belonging to its discrete spectrum. Thus, for $A > 2$ one can write

$$\sigma_0^A = \sigma_0^{A(3q)} \qquad , \qquad \sigma_1^A = P_{3q}^A\ \sigma_1^{A(3q)}/S_1 + P_{6q}^A\ \sigma^{A(6q)} \qquad (7)$$

where $P_{3q}^A + P_{6q}^A = S_1$, with $S_1 \equiv \int dk dE\ P_1^N(k, E)$ being the total probability that, after the removal of a nucleon, the residual $(A-1)$ system is in any state of its continuum. In Eq. (7) the $6q$ bag contribution $\sigma^{A(6q)}$ reads as follows

$$\sigma^{A(6q)} = \frac{A}{2}\ \sigma_{Mott} \sum_\beta \int dW_{6q} d\mathbf{k}_{CM} \delta\left[\nu + k_{CM}^0 - \sqrt{W_{6q}^2 + (\mathbf{k}_{CM} + \mathbf{q})^2}\right] \cdot$$

$$\frac{W_{6q}}{\sqrt{W_{6q}^2 + (\mathbf{k}_{CM} + \mathbf{q})^2}}\ n_{CM}^\beta(\mathbf{k}_{CM}) \sum_{j=1,2} b_j^{(\beta)}\ W_j^{(\beta)}(W_{6q}, Q^2) \quad (8)$$

where $\beta = (u^2d^4, u^3d^3, u^4d^2) = ([nn], [np], [pp])$ identifies the type of $6q$ bag; $k_{CM}^0 \equiv M_A - \sqrt{M_{A-2}^2 + k_{CM}^2}$ is the initial $6q$ bag energy in the lab system; $n_{CM}^\beta$ is the $CM$ momentum distribution of the $6q$ bag. For $A > 2$ the effects from the Fermi motion of the $CM$ of the $6q$ bag can be estimated by adopting the extended $2NC$ model, i.e. by using for the $CM$ momentum distribution of a $6q$ bag the one of a correlated $NN$ pair (i.e., $n_{CM}^\beta = n_{CM}^{N_1N_2}$); this implies that the introduction of a $6q$ bag is assumed to modify only the intrinsic structure of a $NN$ cluster at short separations. Since in Eq. (8) only low-momentum components (i.e., $k_{CM} < 1.5\ fm^{-1}$) have to be considered in $n_{CM}^{N_1N_2}$ (cf. [1]), the coefficients $b_j^{(\beta)}$ can be safely taken equal to $b_1^{(\beta)} = 2tg^2(\theta_e/2)$ and $b_2^{(\beta)} = 1$.

Within the extended $2NC$ model the inclusive process $A(e, e')X$ can be investigated for any value of $A$, including the contributions both from $3q$ (Eq. (4)) and $6q$ (Eq. (8)) cluster configurations. The results of the calculations, performed for the processes $^2H(e, e')X$ and $^4He(e, e')X$ at $Q^2 \sim 15\ (GeV/c)^2$, are reported in Fig. 1. It can be seen that: i) up to $x \sim 1.5$ the $DIS$ contribution, due to virtual photon absorption by a quark belonging to a nucleon in the nucleus, overwhelms the contamination due to quasi-elastic $(QE)$ scattering processes; ii) the kinematical regions corresponding to $x > 1.5$ appear to be appropriate for investigating the effects of $6q$ cluster configurations in light nuclei, provided the value of $P_{6q}^A$ is sufficiently large.

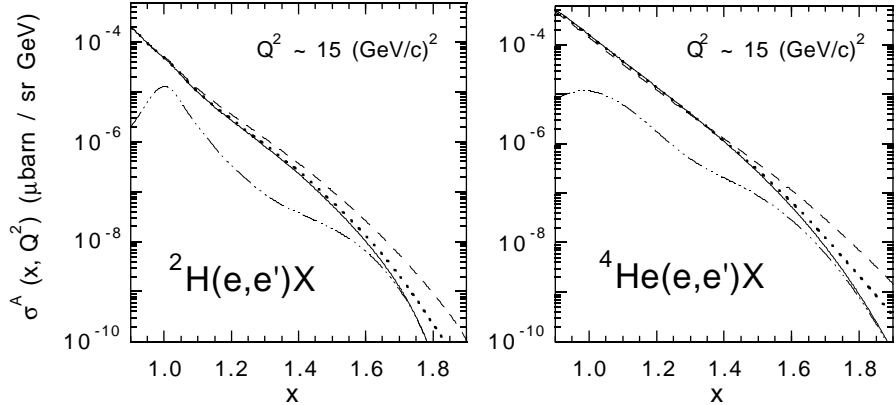

**Figure 1.** The cross section for the inclusive processes $^2H(e,e')X$ and $^4He(e,e')X$ versus $x$ at $Q^2 \sim 15\ (GeV/c)^2$. The dash-dotted lines are the $QE$ contribution, evaluated as in [3], whereas the solid lines include the $DIS$ contribution calculated using in Eq. (4) the free nucleon structure function of Ref. [2]. The dotted (dashed) lines are the results obtained assuming $P_{6q}^{^2H} = 1\%\ (5\%)$ in Eq. (5) and $P_{6q}^{^4He} = 2\%\ (10\%)$ in Eq. (7) for the probability of a $6q$ bag admixture in $^2H$ and $^4He$.

It should be therefore desirable to find an observable, which does not depend crucially upon the value of $P_{6q}^A$, not known from quark counting rules. To this end, the energy spectra of nucleons emitted in semi-inclusive $DIS$ processes $A(\ell, \ell'N)X$ off nuclei, in which, besides the scattered lepton, a nucleon is detected in the final state, could be considered. The relevant nucleon production mechanisms in the process $A(\ell, \ell'N)X$ have been analyzed for $A > 2$ in Ref. [5], where it has been shown that both at $x < 1$ and $x > 1$ the emission of nucleons in the forward hemisphere appears to be the most appropriate kinematical condition for studying multiquark cluster configurations in nuclei, provided the energy distribution of the emitted nucleons is investigated at high energy ($> 0.5\ GeV$). However, at $x > 1$, in case of inclusive as well as semi-inclusive processes, the $DIS$ contribution dominates over the $QE$ scattering one only at very large values of $Q^2$ ($> 15\ (GeV/c)^2$) (cf. [3] and [5]). Therefore, it is worth noting that in case of a deuteron target the production of nucleons arising from $QE$ scattering processes can be disentangled kinematically from the nucleon emission due to $DIS$ events. As a matter of fact, energy and momentum conservations imply that $\nu + M_D = \sqrt{M^2 + p^2} + \sqrt{M_X^2 + (\boldsymbol{q} - \boldsymbol{p})^2}$, where $\nu$ ($\boldsymbol{q}$) is the energy (three-momentum) transfer, $M_D$ is the deuteron mass, $M_X$ is the invariant mass of the residual system and $\boldsymbol{p}$ is the momentum of the detected nucleon. It follows that for fixed values of $\nu$, $\boldsymbol{q}$ and $\boldsymbol{p}$ the value of $M_X$ is kinematically known. Therefore, it is possible to distinguish $QE$ scattering events ($M_X = M$) from $DIS$ ones ($M_X > M$), so that the $DIS$ contribution to the semi-inclusive process $^2H(e, e'N)X$ can be investigated at moderate values of $Q^2$ ($\sim 5 \div 10\ (GeV/c)^2$). A sample of the results, obtained by applying the approach of Ref. [5] to the reaction $^2H(e,e'p)X$, is reported in Fig. 2. It can be clearly seen that at $x > 1$ the slope of the energy distribution of forward emitted protons is almost independent of $P_{6q}^{^2H}$, but still sharply sensitive to $6q$ bag effects, provided sufficiently high values of the kinetic energy of the detected proton are considered.

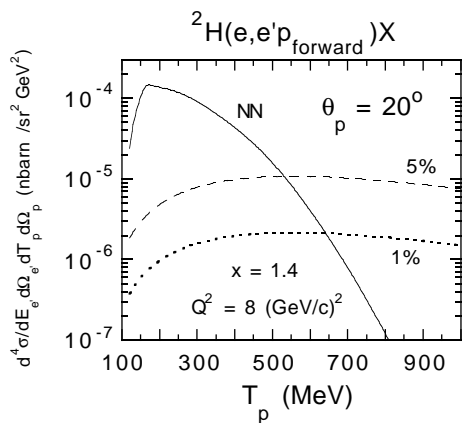

**Figure 2.** The $DIS$ cross section for the semi-inclusive process $^2H(e,e'p)X$ versus the kinetic energy $T_p$ of the detected proton, emitted forward at $\theta_p = 20°$, when $x = 1.4$ and $Q^2 = 8\ (GeV/c)^2$. The solid line is the $DIS$ contribution due to virtual photon absorption on a $NN$ pair, whereas the dotted (dashed) line corresponds to virtual photon absorption by a quark belonging to a $6q$ bag only, assuming $P_{6q}^{^2H} = 1\%\ (5\%)$.

In conclusion, inclusive $A(e,e')X$ and semi-inclusive $A(e,e'N)X$ processes off few-nucleon systems have been investigated at $x > 1$, showing some of the relevant features of the cross section, which are sensitive to the effects arising from nucleon-nucleon correlations as well as from the possible presence of exotic multiquark cluster configurations at short internucleon separations.